\documentclass[12pt,a4paper]{article}

\usepackage{amsmath}

\usepackage{amssymb}

\usepackage{amsfonts}

\usepackage{epsfig}

\usepackage{exscale}

\usepackage{float}

\usepackage{bbm}

\usepackage[numbers,sort&compress]{natbib}

\usepackage{tikz}

\newcommand{\Z}{{\mathbb{Z}}}

\newcommand{\R}{{\mathbb{R}}}

\newcommand{{\1}}{{\mathbbm{1}}}

\makeatletter

\@addtoreset{equation}{section}

\makeatother

\title{Rotor Spectra and Berry Phases in the \\
Chiral Limit of QCD on a Torus}

\author{N.\ D.\ Vlasii and U.-J.\ Wiese \\ \\
{\small Albert Einstein Center for Fundamental Physics,} \\ 
{\small Institute for Theoretical Physics,} \\
{\small Bern University, Sidlerstrasse 5, CH-3012 Bern, Switzerland} \\ \\}

\begin{document} 

\maketitle

\begin{abstract} \normalsize

We consider the finite-volume spectra of QCD in the chiral limit of massless
up and down quarks and massive strange quarks in the baryon number sectors
$B = 0$ and $B = 1$ for different values of the isospin. Spontaneous symmetry 
breaking gives rise to rotor spectra, as the chiral order parameter precesses 
through the vacuum manifold. Baryons of different isospin influence the motion 
of the order parameter through non-trivial Berry phases and associated abstract 
monopole fields. Our investigation provides detailed insights into the dynamics 
of spontaneous chiral symmetry breaking in QCD on a torus. It also sheds new 
light on Berry phases in the context of quantum field theory. Interestingly, 
the Berry gauge field resulting from QCD solves a Yang-Mills-Chern-Simons 
equation of motion on the vacuum manifold $SU(2) = S^3$.

\end{abstract}

\newpage

\section{Introduction}

Nowadays lattice QCD calculations are performed close to the physical point, 
i.e.\ with realistic quark masses. Very interesting effects also arise 
in the chiral limit of exactly massless up and down quarks, where lattice QCD 
doesn't work so efficiently. In this paper, we focus our theoretical study on 
this somewhat academic limit in order to gain a deeper
understanding of the dynamics of spontaneous chiral symmetry breaking in a 
finite volume. In the absence of any explicit breaking of chiral symmetry, in 
the infinite volume QCD has infinitely many degenerate vacuum states, among 
which one is selected spontaneously. In a finite volume, on the other hand, all
vacuum states coexist, but they are no longer exactly degenerate. Instead their 
energies split into a rotor spectrum with energy differences that are inversely
proportional to the spatial volume. The rotor spectrum in the vacuum sector of 
QCD in the chiral limit was first derived by Leutwyler in \cite{Leu87}. The
dynamics of the chiral order parameter then reduces to the quantum mechanical
motion of a rotor in the vacuum manifold. The rotor spectrum in the baryon 
number $B = 1$ and isospin $I = \frac{1}{2}$ sector has been investigated in 
\cite{Cha08}. The presence of a nucleon influences the precession of the chiral 
order parameter and leaves an imprint on the corresponding rotor spectrum. This
manifests itself by a non-Abelian Berry phase and an abstract monopole field in
the vacuum manifold. In this way, the Berry phase \cite{Ber84,Sim83}, which is
familiar from quantum mechanics, arises even in the context of QCD. It should 
be noted that the nucleon mass, which originates from spontaneous chiral
symmetry breaking, remains non-zero in the chiral limit in a finite volume.
This is the case even when all vacuum states are sampled by the precessing 
order parameter, such that, at least in a naive sense, chiral symmetry is no 
longer spontaneously broken. In this paper, we extend the previous studies to 
baryon sectors with general isospin $I$. Again, non-trivial Berry phases and 
monopole fields explain the resulting rotor spectra. 

Similar situations also arise in $(2+1)$-d antiferromagnets of finite volume 
with a spontaneously broken $SU(2)_s$ spin symmetry. Considering a quadratic 
periodic volume, the quantum mechanical rotor spectrum of the precessing 
staggered magnetization order parameter was first calculated by Hasenfratz and 
Niedermayer in \cite{Has93}. While the analog of the Goldstone pions in QCD are 
massless spinwaves (or magnons) in an antiferromagnet, the condensed matter 
analog of protons and neutrons are holes or electrons doped into an 
antiferromagnet. In this case, again Berry phases and corresponding monopole 
fields describe how a doped hole or electron influences the rotor spectrum 
associated with the precessing staggered magnetization \cite{Cha08}. 
Interestingly, the Berry gauge field that arises in this case is the same as
the one associated with the rotation of diatomic molecules 
\cite{Her63,Mea80,Moo86}. In this paper, we concentrate entirely on QCD, and 
leave the investigation of condensed matter analogs for future study.

In QCD we encounter a Berry gauge field that is defined on the group manifold
of the $SU(2)$ flavor group. Remarkably, while the Berry gauge field is just a
geometric object, it turns out to be a classical solution of a 
Yang-Mills-Chern-Simons equation of motion. The covariantly conserved 
Chern-Simons current then provides a source for the non-Abelian Berry field 
strength. This is similar to the problem addressed in \cite{Son09}. In that 
case, the Berry gauge field corresponds to a Bogomolnyi-Prasad-Sommerfield (BPS)
monopole solution of a Yang-Mills-Higgs equation of motion. There an adjoint
Higgs field, which plays the role of a ``Berry matter field'', gives rise to a
conserved current. 

The rest of this paper is organized as follows. Section 2 summarizes baryon
chiral perturbation theory for non-relativistic baryons. In order to make the
paper self-contained, Sections 3 and 4 summarize the derivation of the rotor
spectrum in the vacuum (i.e.\ for baryon number $B = 0$) and in the nucleon
sector (with $B = 1$ and isospin $I = \frac{1}{2}$), respectively. We also 
provide further details that go beyond the presentation in \cite{Cha08}.
Section 5 provides the major new results of this work, by addressing baryon 
sectors of general isospin. In Section 6 we investigate the nature of the 
Berry gauge field, in particular, as a solution of a classical 
Yang-Mills-Chern-Simons equation of motion on the $SU(2)$ group manifold.
Finally, Section 7 contains our conclusions.

\section{Baryon Chiral Perturbation Theory}

In this section we review baryon chiral perturbation theory for 
non-relativistic baryons \cite{Gas88,Jen91,Jen92,Ber92}, which is an 
appropriate tool to address the low-energy physics of the precessing chiral 
order parameter in the presence of a baryon. We consider the chiral limit of 
massless up and down quarks ($m_u = m_d = 0$), but we treat the strange 
quark as massive ($m_s \neq 0$). The chiral symmetry group of QCD then is
$G = SU(2)_L \times SU(2)_R \times U(1)_B$, which is spontaneously broken to the
subgroup $H = SU(2)_{L=R} \times U(1)_B$, where $SU(2)_{L=R}$ and $U(1)_B$ are the
unbroken isospin and baryon number symmetries, respectively. As a consequence of
chiral symmetry breaking, there are three massless Goldstone pion fields 
$\pi^a(x)$ which
are described by 
\begin{equation}
\label{Uequation}
U(x) = \exp[i \pi^a(x) \tau_a/F_\pi] \in G/H = SU(2) \ ,
\end{equation}
where
$x = (\vec x,t)$ is a point in Euclidean space-time, $F_\pi$ is the pion decay 
constant, and the Pauli matrices $\tau_a$  are the generators of $SU(2)$. 
The spatial zero-mode of $U(x)$ describes the orientation of the chiral order 
parameter in the vacuum manifold $SU(2)$, while non-zero modes correspond to 
pion excitations.

Under global chiral rotations, the field $U(x)$ transforms as
\begin{equation}
U(x)'= L U(x) R^{\dagger} \ , \quad L \in SU(2)_L \ , \ R \in SU(2)_R \ ,
\end{equation}
while under charge conjugation $C$ and parity $P$ it transforms as
\begin{equation}
^C U(x) = U(x)^{\sf T} \ , \quad ^P U(\vec x,t) = U(- \vec x,t)^\dagger \ . 
\end{equation}

In order to couple the chiral order parameter field $U(x)$ to baryon fields, we
first introduce the field $u(x) = U(x)^{1/2} \in SU(2)$. In order to fix the 
sign ambiguity of the square-root, it is important to take $U(x)^{1/2}$ at the 
midpoint of the shortest geodesic in the $SU(2)$ group manifold that connects 
$U(x)$ with the unit-element ${\1}$ \cite{Cha03}. Next, we use $u(x)$ to 
construct a non-Abelian $SU(2)$ gauge field $v_\mu(x)$ and a ``charged'' 
axial-vector field $a_\mu(x)$ as
\begin{eqnarray}
\label{vaequation}
&&v_\mu(x) = \frac{1}{2} [u(x)^\dagger \partial_\mu u(x) + 
u(x) \partial_\mu u(x)^\dagger] = i v_\mu^a(x) \tau_a \ , \nonumber \\
&&a_\mu(x) = \frac{i}{2} [u(x)^\dagger \partial_\mu u(x) - 
u(x) \partial_\mu u(x)^\dagger] = a_\mu^a(x) \tau_a \ .
\end{eqnarray}
Under chiral rotations these fields transforms as
\begin{eqnarray}
&&u(x)' = L u(x) V(x)^\dagger = V(x) u(x) R^\dagger \ , \nonumber \\
&&v_\mu(x)' = V(x) [v_\mu(x) + \partial_\mu] V(x)^\dagger \ , \nonumber \\
&&a_\mu(x)' = V(x) a_\mu(x) V(x)^\dagger \ .
\end{eqnarray}
Interestingly, the global chiral rotations $L$ and $R$ give rise to the local
transformation
\begin{equation}
\label{Vequation}
V(x) = R[R^\dagger L U(x)]^{1/2} [U(x)^{1/2}]^\dagger = 
L[L^\dagger R U(x)^\dagger]^{1/2} U(x)^{1/2} \ ,
\end{equation}
for which $v_\mu(x)$ acts as a composite $SU(2)$ isospin ``gauge'' field.

Under charge conjugation and parity the fields $u(x)$, $v_\mu(x)$, and $a_\mu(x)$
transform as
\begin{eqnarray}
\label{vaCP}
&&^C u(x) = u(x)^{\sf T} \ , \qquad ^P u(\vec x,t) = u(- \vec x,t)^\dagger \ , 
\nonumber\\
&&^C v_\mu(x) = v_\mu(x)^* \ , \qquad ^P v_i(\vec x,t) = - v_i(- \vec x,t) \ , 
\quad ^P v_t(\vec x,t) = v_t(- \vec x,t) \ , \nonumber\\
&&^C a_\mu(x) = a_\mu(x)^* \ , \qquad ^P a_i(\vec x,t) = a_i(- \vec x,t) \ , 
\quad ^P a_t(\vec x,t) = - a_t(- \vec x,t) \ .
\end{eqnarray}
Here the index $i$ denotes a spatial direction and $t$ denotes Euclidean time.

In addition to the chiral order parameter field $U(x)$ we consider 
non-relativistic baryon fields $\Psi(x)$ and $\Psi^\dagger(x)$ which are 
2-component Pauli spinors associated with spin $\frac{1}{2}$. In QCD these form
the baryon octet consisting of the nucleons $N$, as well as the $\Sigma$, 
$\Lambda$, and $\Xi$ baryons, with isospin $\frac{1}{2}$, 1, 0, and 
$\frac{1}{2}$, respectively. Under parity, the non-relativistic baryon fields 
transform as
\begin{equation}
^P \Psi(\vec x,t) = \Psi(- \vec x,t) \ , \quad 
^P \Psi^\dagger(\vec x,t) = \Psi^\dagger(- \vec x,t) \ .
\end{equation}
Under charge conjugation, these fields would transform into anti-baryon fields
which we are not considering in our non-relativistic approach.

Until now we have constructed the $SU(2)$ matrix $u(x)$ and the corresponding
algebra-valued fields $v_\mu(x)$ and $a_\mu(x)$ that act on the 
$I = \frac{1}{2}$ representation. These can be used to couple the chiral order
parameter to nucleon and $\Xi$ baryon fields. The nucleon field is a 
2-component isospinor
\begin{equation}
N(x) = \left(\begin{array}{c} p(x) \\ n(x) \end{array}\right) \ , \quad
N^\dagger(x) = (p^\dagger(x), n^\dagger(x)) \ ,
\end{equation}
consisting of proton and neutron fields $p(x)$ and $n(x)$, which transforms 
under chiral rotations as
\begin{equation}
N(x)' = V(x) N(x) \ , \quad N^\dagger(x)' = N^\dagger(x) V(x)^\dagger \ .
\end{equation}
Chirally invariant terms can then be constructed with covariant derivatives
\begin{equation}
D_\mu N(x) = (\partial_\mu + v_\mu(x)) N(x) \ , \quad 
D_\mu N(x)' = V(x) D_\mu N(x) \ .
\end{equation}
Since $\Xi$ baryons also have isospin $\frac{1}{2}$, they transform in the same 
way as the nucleon fields.

$\Sigma$ baryons, on the other hand, have isospin 1 and can be represented as
\begin{eqnarray}
&&\Sigma(x) = \left(\begin{array}{cc} \Sigma_0(x) & \sqrt{2} \Sigma_+(x) \\
\sqrt{2} \Sigma_-(x)  & - \Sigma_0(x) \end{array} \right) \ , \nonumber \\
&&\Sigma^\dagger(x) = \left(\begin{array}{cc}
\Sigma_0^\dagger(x) & \sqrt{2} \Sigma_-^\dagger(x)  \\
\sqrt{2} \Sigma_+^\dagger(x) & - \Sigma_0^\dagger(x) \end{array} \right) \ .
\end{eqnarray}
They transform in the adjoint representation of $SU(2)$ isospin, i.e.
\begin{equation}
\Sigma(x)' = V(x) \Sigma(x) V(x)^\dagger \ ,
\end{equation}
and the corresponding covariant derivative takes the form
\begin{equation}
\label{covariantAdjRep}
D_\mu \Sigma(x) = \partial_\mu \Sigma(x) + [v_\mu(x),\Sigma(x)] \ , \quad
D_\mu \Sigma(x)' = V(x) \Sigma(x) V(x)^\dagger \ .
\end{equation}

In order to facilitate a generalization to baryons with arbitrary isospin,
let us now rewrite the $\Sigma$ baryon field explicitly as an isovector 
$\vec \Sigma(x)$ such that
\begin{eqnarray}
&&\Sigma(x) = \vec \Sigma(x) \cdot \vec \tau = 
(\Sigma_1(x),\Sigma_2(x),\Sigma_3(x)) \cdot \vec \tau \ , \nonumber \\
&&\Sigma_\pm(x) = \frac{1}{\sqrt{2}} (\Sigma_1(x) \mp i \Sigma_2(x)) \ , \
\Sigma_0(x) = \Sigma_3(x) \ .
\end{eqnarray}
In this representation, the covariant derivative takes the form
\begin{equation}
D_\mu \vec \Sigma(x) = 
\partial_\mu \vec \Sigma(x) + \vec v_\mu(x) \times \vec \Sigma(x) \ ,
\end{equation}
and the field transforms under global chiral rotations as
\begin{equation}
\label{Sigmatransform}
\vec \Sigma(x)' = {\cal V}(x) \vec \Sigma(x) \ , \quad
D_\mu \vec \Sigma(x)' = {\cal V}(x) D_\mu \vec \Sigma(x) \ .
\end{equation}
Here ${\cal V}(x)$ is an orthogonal $3 \times 3$ matrix that is related to 
$V(x)$ (cf.\ eq.(\ref{Vequation})) by
\begin{equation}
{\cal V}_{ab}(x) = \frac{1}{2} \mbox{Tr} [V(x)^\dagger \tau_a V(x) \tau_b] \ ,
\end{equation}
i.e.\ ${\cal V}(x) \in SO(3)$ is the adjoint representation version of the 
original transformation $V(x) \in SU(2)$ that acts in the fundamental 
representation. 

Because it is mathematically feasible, we will now consider baryons with 
arbitrary isospin, even if they don't exist in QCD. We generalize the 
construction to baryon fields of arbitrary isospin $I$ with the generators 
$T_a$ of the $(2I+1)$-dimensional representation. First we introduce the 
Goldstone boson field
\begin{equation}
\label{Oequation}
O(x) = \exp[2 i \pi^a T_a/F_\pi] \ ,
\end{equation}
which replaces the field $U(x) \in SU(2)$ of eq.(\ref{Uequation}) that acts in
the fundamental representation for which $T_a = \frac{1}{2} \tau_a$. For the
adjoint $I = 1$ representation, one then obtains
\begin{equation}
O_{ab}(x) = \frac{1}{2} \mbox{Tr} [U(x)^{\dagger} \tau_a U(x) \tau_b] \ .
\end{equation}
We now use the parametrization
\begin{equation}
U(x) = \exp[i \vec \pi(x) \cdot \vec \tau/F_\pi] = 
\cos \alpha(x) \1 + 
i \sin\alpha(x) \vec e_\alpha(x) \cdot \vec \tau \ ,
\end{equation}
where $\alpha(x) = |\vec \pi(x)|/F_\pi$ and $\vec e_\alpha(x)$ is the unit-vector
pointing in the direction of $\vec \pi(x)$. The map from the fundamental to 
the adjoint representation can also be expressed as
\begin{equation}
O(x) = \exp[2 i \vec \pi(x) \cdot \vec T/F_\pi] =
\1 + 2 i \cos\alpha(x) \sin \alpha(x) \vec e_\alpha(x) \cdot \vec T 
- 2 \sin^2\alpha(x) (\vec e_\alpha(x) \cdot \vec T)^2 \ .
\end{equation}
For isospin $I = \frac{3}{2}$ (with $(4 \times 4)$-matrix 
generators $T_a$) one obtains
\begin{eqnarray}
O(x)&=&\left(\frac{3}{2} - \frac{1}{2} \cos^2\alpha(x)\right)\cos\alpha(x) \1 +
i \left(2 + \frac{1}{3} \sin^2\alpha(x)\right) \sin\alpha(x) 
\vec e_\alpha(x) \cdot \vec T \nonumber \\
&-&2 \cos\alpha(x) \sin^2\alpha(x) (\vec e_\alpha(x) \cdot \vec T)^2 -
\frac{4}{3} i \sin^3\alpha(x) (\vec e_\alpha(x) \cdot \vec T)^3 \ .
\end{eqnarray}
Similar expressions exist for higher values of the isospin.

For general isospin, under global chiral rotations the field $O(x)$ transforms 
as
\begin{equation}
O(x)'= {\cal L} O(x) {\cal R}^\dagger \ .
\end{equation}
As for the fundamental representation, we now define $o(x) = O(x)^{1/2}$, which
transforms as
\begin{equation}
o(x)' = {\cal L} o(x) {\cal V}(x)^\dagger = {\cal V}(x) o(x) {\cal R}^\dagger \ ,
\end{equation}
where the matrix ${\cal V}(x)$ is given by
\begin{equation}
{\cal V}(x)= {\cal R}[{\cal R}^\dagger {\cal L} O(x)]^{1/2}[O(x)^{1/2}]^\dagger =
{\cal L}[{\cal L}^\dagger {\cal R} O(x)^\dagger]^{1/2} O(x)^{1/2} \ .
\end{equation}
In analogy to eq.(\ref{vaequation}), the corresponding vector and axial-vector
fields now take the form
\begin{eqnarray}
\label{nualpha}
&&\nu_\mu(x) = \frac{1}{2} 
[o(x)^\dagger \partial_\mu o(x) + o(x) \partial_\mu o(x)^\dagger] = 
2 i v_\mu^a(x) T_a \ , \nonumber \\
&&\alpha_\mu(x) = 
\frac{i}{2} [o(x)^\dagger \partial_\mu o(x) - o(x) \partial_\mu o(x)^\dagger] =
2 a_\mu^a(x) T_a \ ,
\end{eqnarray}
which under the global chiral rotations transform as
\begin{eqnarray}
&&\nu_\mu(x)' = {\cal V}(x)[\nu_\mu(x) + \partial_\mu]{\cal V}(x)^\dagger \ , 
\nonumber \\
&&\alpha_\mu(x)' = {\cal V}(x) \alpha_\mu(x) {\cal V}(x)^\dagger \ .
\end{eqnarray}
It is important to point out that $v_\mu^a(x)$ and $a_\mu^a(x)$ in 
eq.(\ref{nualpha}) are the same fields as in eq.(\ref{vaequation}), independent 
of the representation, which enters $\nu_\mu(x)$ and $\alpha_\mu(x)$ only via 
the generators $T_a$. Under charge conjugation $C$ and parity $P$ the fields 
$\nu_\mu(x)$ and $\alpha_\mu(x)$ transform exactly like $v_\mu(x)$ and $a_\mu(x)$ 
(cf.\ eq.(\ref{vaCP})).

A baryon field $\Psi(x)$ with general isospin $I$ (i.e.\ a $(2I+1)$-dimensional
isospinor with spin $\frac{1}{2}$) then transforms just like $\vec \Sigma(x)$ 
for $I = 1$ (cf.\ eq.(\ref{Sigmatransform})), i.e.
\begin{equation}
\Psi(x)' = {\cal V}(x) \Psi(x) \ , \quad 
\Psi^\dagger(x)' = \Psi^\dagger(x) {\cal V}(x) \ ,
\end{equation}
and the corresponding covariant derivative is given by
\begin{equation}
D_\mu \Psi(x) = (\partial_\mu + \nu_\mu(x)) \Psi(x) \ , \quad 
D_\mu \Psi(x)' = {\cal V}(x) D_\mu \Psi(x) \ .
\end{equation}

\section{Rotor Spectrum in the $B = 0$ Vacuum Sector}

In this section we consider the QCD spectrum in a periodic spatial volume 
$V = L^3$ in the vacuum sector, i.e.\ for baryon number $B = 0$. In the chiral
limit of massless up and down quarks, as a consequence of spontaneously broken
$SU(2)_L \times SU(2)_R$ chiral symmetry, there are infinitely many exactly
degenerate ground states, at least in an infinite volume. In a finite periodic
volume, the energies of these states split into a rotor spectrum, which was
first derived in \cite{Leu87} in the so-called $\delta$-regime of chiral 
perturbation theory \cite{Gas85}. As a preparation for the $B \neq 0$ case, 
and in order to make this paper self-contained, here we review the $B = 0$ 
case.

The low-energy dynamics of the chiral order parameter field 
$U(\vec x,t) \in SU(2)$ is governed by the Euclidean action
\begin{equation}
S[U] = \int_V d^3x \int dt \ \frac{F_\pi^2}{4}  
\mbox{Tr}[\partial_\mu U^\dagger \partial_\mu U] \ .
\end{equation} 
Since we consider the chiral limit in a finite volume, we are in the
$\delta$-regime of chiral perturbation theory, in which the dynamics are 
dominated by the spatially-independent zero-mode $U(t)$ of the order parameter 
field. After integrating out the spatial non-zero modes of $U(\vec x,t)$, the 
dynamics reduces to the quantum mechanical motion of the zero-mode $U(t)$ and 
one obtains 
\begin{equation}
S[U] = \int dt \ \frac{\Theta}{4} \mbox{Tr}[\partial_t U^\dagger \partial_t U] 
\ .
\end{equation}
Here $\Theta$ is the moment of inertia of a quantum rotor that precesses in the 
vacuum manifold $SU(2)$. At tree level, it takes the value 
$\Theta = F_\pi^2 L^3$. Higher-order 1- and 2-loop corrections were worked out 
in \cite{Has10,Nie11} for the $O(N)$ model in $(2+1)$ and $(3+1)$ dimensions.
They also apply to $(3+1)$-d QCD with two flavors because the chiral symmetry 
group is then given by $SU(2)_L \times SU(2)_R = O(4)$. 

Parametrizing the 3-sphere $S^3 = SU(2)$ as
\begin{eqnarray}
\label{parametrization}
&&U(t) = \cos\alpha(t) + i \sin\alpha(t) \vec e_\alpha(t) \cdot \vec{\tau} \ ,
\nonumber \\
&&\vec e_\alpha(t) = \left(\sin\theta(t) \cos\varphi(t), 
\sin\theta(t) \sin\varphi(t),\cos\theta(t)\right) \ ,
\nonumber \\
&&\vec e_\theta(t) = \left(\cos\theta(t) \cos\varphi(t),
\cos\theta(t) \sin\varphi(t),- \sin\theta(t)\right) \ ,
\nonumber \\
&&\vec e_\varphi(t) = \left(- \sin\varphi(t),\cos\varphi(t),0\right) \ ,
\end{eqnarray}
the corresponding Lagrange function takes the form
\begin{equation}
L(\alpha,\partial_t\alpha;\theta,\partial_t\theta;\varphi,\partial_t \varphi)
= \frac{\Theta}{2} \left[(\partial_t \alpha)^2 + \sin^2\alpha 
\left((\partial_t \theta)^2 + \sin^2\theta (\partial_t \varphi)^2\right)\right]
\ .
\end{equation}
The momenta canonically conjugate to $\alpha$, $\theta$, and $\varphi$ then are
\begin{eqnarray}
&&p_\alpha = \frac{\delta L}{\delta\partial_t \alpha} =
\Theta \partial_t \alpha \ , \nonumber\\
&&p_\theta = \frac{\delta L}{\delta\partial_t \theta} =
\Theta \sin^2\alpha \partial_t \theta \ , \nonumber\\
&&p_\varphi = \frac{\delta L}{\delta\partial_t \varphi} =
\Theta \sin^2\alpha \sin^2\theta \partial_t \varphi \ .
\end{eqnarray}
After canonical quantization, the resulting Hamilton operator is the Laplacian 
on the sphere $S^3$ 
\begin{equation}
H = - \frac{1}{2 \Theta} \left\{\frac{1}{\sin^2\alpha} \partial_\alpha
[\sin^2\alpha \partial_\alpha] + \frac{1}{\sin^2\alpha \sin\theta}
\partial_\theta [\sin\theta \partial_\theta] +
\frac{1}{\sin^2\alpha \sin^2\theta} \partial_\varphi^2\right\} \ . 
\end{equation}
In terms of the $SU(2)_L$ and $SU(2)_R$ generators, this is equal to
\begin{equation}
\label{hamiltonian_vacuum}
H = \frac{1}{\Theta} \left(\vec L^2 + \vec R^2\right) =
\frac{1}{2 \Theta} \left(\vec J^2 + \vec K^2\right) \ ,
\end{equation}
where 
\begin{eqnarray}
\label{left_right_0}
&&\vec L = \frac{1}{2}(\vec J - \vec K) \ , \qquad
\vec R = \frac{1}{2}(\vec J + \vec K) \ , \nonumber \\
&&J_\pm = \exp(\pm i \varphi)(\pm \partial_\theta + i \cot\theta \partial_\varphi)
\ ,
\nonumber\\
&&J_3 = - i \partial_\varphi \ ,\nonumber \\
&&K_\pm = \exp(\pm i \varphi) \left(i \sin\theta \partial_\alpha + 
i \cot\alpha \cos\theta \partial_\theta \mp \frac{\cot\alpha}{\sin\theta}
\partial_\varphi \right) \ , \nonumber \\
&&K_3 = i \left(\cos\theta \partial_\alpha - \cos\alpha \sin\theta 
\partial_\theta \right) \ .
\end{eqnarray}
Since $SU(2)_L \times SU(2)_R$ has rank 2, there are two Casimir operators,
$\vec L^2$ and $\vec R^2$, or alternatively 
\begin{equation}
C_1 = \vec R^2 + \vec L^2 = \frac{1}{2} (\vec J^2 + \vec K^2) \ , \qquad
C_2 = \vec R^2 - \vec L^2 = \vec J \cdot \vec K \ .
\end{equation}
The Casimir operator $C_1$ determines the spectrum of the Hamiltonian
eq.(\ref{hamiltonian_vacuum}) as
\begin{equation}
E_{j_L,j_R} = \frac{1}{\Theta} [j_L(j_L+1) + j_R(j_R+1)] \ ,
\end{equation}
where $j_L$ and $j_R$ are integer or half-integer. It is important to note that
not all combinations of $j_L$ and $j_R$ are allowed. Using the explicit
expressions for $\vec J$ and $\vec K$ of eq.(\ref{left_right_0}), it is 
straightforward to show that $C_2 = \vec J \cdot \vec K = 0$, which implies
$j_L = j_R$. Introducing $j = j_L + j_R \in \{0,1,2,\dots\}$, the energy 
spectrum then takes the form
\begin{equation}
E_j = \frac{j(j+2)}{2 \Theta} \ ,
\end{equation}
and each state is $(2 j_L + 1)(2 j_R + 1) = (j+1)^2$-fold degenerate. The
scaling of the energy $E_j$ with the Casimir operator eigenvalue
$C_1 = \frac{1}{2} j(j + 2)$ persists even at the 2-loop level of chiral
perturbation theory \cite{Has10,Nie11}. Tiny corrections proportional to 
$C_1^2$, which arise at the 3-loop level, were identified in \cite{Nie18}.
Explicit chiral symmetry breaking effects due to non-zero up and down
quark masses have been discussed in \cite{Wei10}. Low-temperature 
effects in the $\delta$-regime as well as the transition to the 
$\varepsilon$-regime were considered in \cite{Mat16}.

It is interesting to confront the analytic results for $E_j$ with lattice
QCD Monte Carlo data. Performing lattice QCD simulations directly in the
chiral limit is very challenging, because the standard hybrid Monte Carlo 
algorithm then no longer works efficiently. Still, partly by extrapolating 
lattice data from the p-regime into the $\delta$-regime, reasonable 
agreement with the analytic results has been obtained in 
\cite{Has06,Bie10,Bie11}.

\section{Rotor Spectrum in the Presence of a Nucleon}

In this section we consider the effect of a nucleon on the rotor spectrum, i.e.\
we consider the baryon number $B = 1$, isospin $I = \frac{1}{2}$ sector. This
was first investigated in \cite{Cha08}. As a preparation for the case of general
isospin $I$, which will be discussed in Section 5, here we review the 
derivation for $I = \frac{1}{2}$ and add further details that were not discussed
explicitly in \cite{Cha08}. The treatment is based on baryon chiral perturbation
theory for non-relativistic baryons \cite{Gas88,Jen91,Jen92,Ber92} as outlined 
in Section 2.

When a single nucleon of momentum $\vec p = |\vec p| \vec e_p$ 
($\vec p = 2 \pi/L \vec n$ with $\vec n \in \Z^3$) is added to the system, the 
finite-volume low-energy effective Lagrange function takes the form
\begin{equation}
L = \frac{\Theta}{4} \mbox{Tr}[\partial_t U^\dagger \partial_t U] + N^\dagger
\left[E(\vec p) - i \partial_t - i v_t - i \lambda \vec \sigma \cdot \vec e_p 
a_t \right] N \ .
\end{equation}
Here  $E(\vec p) = M + {\vec p}^2/2M$, $M$ is the nucleon mass, and 
$\lambda = g_A |\vec p|/M$, where $g_A$ is the nucleon's axial-vector coupling.
Proton and neutron are distinguished by a flavor index of the Pauli spinor 
$N(t)$. The spin of the nucleon is $\frac{\vec \sigma}{2}$ and its isospin is 
represented by $\frac{\vec \tau}{2}$.

Then, by applying the parametrization of eq.(\ref{parametrization}) to
eq.(\ref{vaequation}), one obtains
\begin{eqnarray}
\label{vtat}
&&v_t = i \sin^2\frac{\alpha}{2}(\partial_t \theta \vec e_\varphi - 
\sin\theta \partial_t \varphi \vec e_\theta) \cdot \vec \tau \ , \nonumber \\
&&a_t = \left(\frac{\partial_t \alpha}{2} \vec e_\alpha +
\sin\alpha \frac{\partial_t \theta}{2} \vec e_\theta +
\sin\alpha \sin\theta \frac{\partial_t \varphi}{2} \vec e_\varphi \right) \cdot 
\vec \tau \ ,
\end{eqnarray}
and the Lagrange function takes the form
\begin{eqnarray}
L&=&\Theta \left[(\partial_t \alpha)^2 + 
\sin^2\alpha \left((\partial_t \theta)^2 + 
\sin^2\theta(\partial_t \varphi)^2 \right)\right] \nonumber \\
&+&N^\dagger \sin^2\frac{\alpha}{2}(\partial_t \theta \vec e_\varphi - 
\sin\theta \partial_t \varphi \vec e_\theta) \cdot \vec \tau N \nonumber\\
&-&N^\dagger i\lambda (\vec \sigma \cdot \vec e_p) 
\left(\frac{\partial_t \alpha}{2} \vec e_\alpha + \sin\alpha 
\frac{\partial_t \theta}{2} \vec e_\theta 
+ \sin\alpha \sin\theta \frac{\partial_t \varphi}{2} \vec e_\varphi \right)
\cdot \vec \tau N \ .
\end{eqnarray}
The canonically conjugate momenta to $\alpha$, $\theta$, and $\varphi$ are 
given by
\begin{eqnarray}
&&p_\alpha = \frac{\delta L}{\delta \partial_t \alpha} =
\Theta \partial_t \alpha + i A_\alpha \ , \nonumber\\
&&p_\theta = \frac{\delta L}{\delta \partial_t \theta} =
\Theta \sin^2\alpha \partial_t \theta + i A_\theta \ , \nonumber\\
&&p_\varphi = \frac{\delta L}{\delta \partial_t \varphi} =
\Theta \sin^2\alpha \sin^2\theta \partial_t \varphi + i A_\varphi \ ,
\end{eqnarray}
with the anti-Hermitean non-Abelian vector potential given by
\begin{eqnarray}
\label{vector_potential}
&&A_\alpha = i \frac{\lambda}{2} (\vec \sigma \cdot \vec e_p) 
\vec e_\alpha \cdot \vec \tau \ , \nonumber \\
&&A_\theta = i \left(\sin^2\frac{\alpha}{2} \vec e_\varphi +
\frac{\lambda}{2} (\vec \sigma \cdot \vec e_p) \sin\alpha \vec e_\theta \right) 
\cdot \vec \tau \ , \nonumber \\
&&A_\varphi = i \left(-\sin^2\frac{\alpha}{2} \sin\theta \vec e_\theta +
\frac{\lambda}{2} (\vec \sigma \cdot \vec e_p) 
\sin\alpha \sin\theta \vec e_\varphi \right) \cdot \vec \tau \ .
\end{eqnarray}
The corresponding Hamiltonian takes the form
\begin{eqnarray}
H&=&- \frac{1}{2 \Theta} \Bigg\{\frac{1}{\sin^2\alpha}
(\partial_\alpha + A_\alpha)[\sin^2\alpha (\partial_\alpha + A_\alpha)] \nonumber 
\\
&+&\frac{1}{\sin^2\alpha \sin\theta}(\partial_\theta + A_\theta)
[\sin\theta (\partial_\theta + A_\theta)] +
\frac{1}{\sin^2\alpha \sin^2\theta}{(\partial_\varphi + A_\varphi)}^2\Bigg\} \ .   
\end{eqnarray}
The non-Abelian vector potential (\ref{vector_potential}) enters the Hamiltonian
as a Berry connection with the associated field strength
\begin{eqnarray}
&&F_{\alpha\theta} = i \frac{(1 - \lambda^2)}{2} \sin\alpha
\vec e_\varphi \cdot \vec \tau \ , \nonumber \\
&&F_{\theta\varphi} = i \frac{(1 - \lambda^2)}{2} \sin^2\alpha \sin{\theta}
\vec e_\alpha \cdot \vec\tau \ , \nonumber \\
&&F_{\varphi\alpha} = i \frac{(1 - \lambda^2)}{2} \sin\alpha \sin\theta
\vec e_\theta \cdot \vec \tau \ .
\end{eqnarray}
In the presence of the nucleon the generators of the chiral rotations take the 
form
\begin{eqnarray}
\label{generators}
J_\pm&=&\exp(\pm i \varphi)(\pm \partial_\theta + i \cot\theta \partial_\varphi) +
\frac{\tau_\pm}{2} \ , \nonumber \\
J_3&=&- i \partial_\varphi + \frac{\tau_3}{2} \ , \nonumber \\
K_\pm&=&\exp(\pm i \varphi) \Bigg(i \sin\theta \partial_\alpha + 
i \cot\alpha \cos\theta \partial_\theta \mp \frac{\cot\alpha}{\sin\theta}
\partial_\varphi \nonumber \\
&\mp&\frac{i}{2} \tan\frac{\alpha}{2} \vec e_\theta \cdot \vec\tau + 
\frac{1}{2} \tan\frac{\alpha}{2} \cos\theta \vec e_\varphi \cdot \vec\tau\Bigg)
\ , \nonumber \\
K_3&=&i \left(\cos\theta \partial_\alpha - \cos\alpha \sin\theta
\partial_\theta\right) - \frac{1}{2} \tan\frac{\alpha}{2} \sin\theta \vec 
e_\varphi \cdot \vec\tau \ .
\end{eqnarray}

For $\lambda = 0$ the Hamiltonian can then be written as
\begin{equation}
H(0) = \frac{1}{2 \Theta} \left(\vec J^2 + \vec K^2 - \frac{3}{4}\right) + 
E(\vec p) \ .
\end{equation}
As in the vacuum sector, not all representations of $SU(2)_L \times SU(2)_R$ are
actually realized. In the presence of a nucleon, the Casimir operators are 
constrained by
\begin{equation}
C_2^2 - \frac{1}{2} C_1 - \frac{3}{16} = 0 \ .
\end{equation}
This follows directly from the explicit form of $\vec J$ and $\vec K$ in 
eq.(\ref{generators}). This constraint can be satisfied only if 
$j_L = j_R \pm \frac{1}{2}$, which implies that the rotor spectrum takes the 
form
\begin{equation}
E_j(0) = \frac{1}{2\Theta} \left[j(j+2) - \frac{1}{2}\right] + E(\vec p) \ .
\end{equation}
Here $j = j_L + j_R \in \{\frac{1}{2},\frac{3}{2},\dots\}$ and each state is 
$2 (j + \frac{1}{2})(j + \frac{3}{2})$-fold degenerate.

\begin{figure}[tbh]
\begin{center}
\includegraphics[width=\columnwidth]{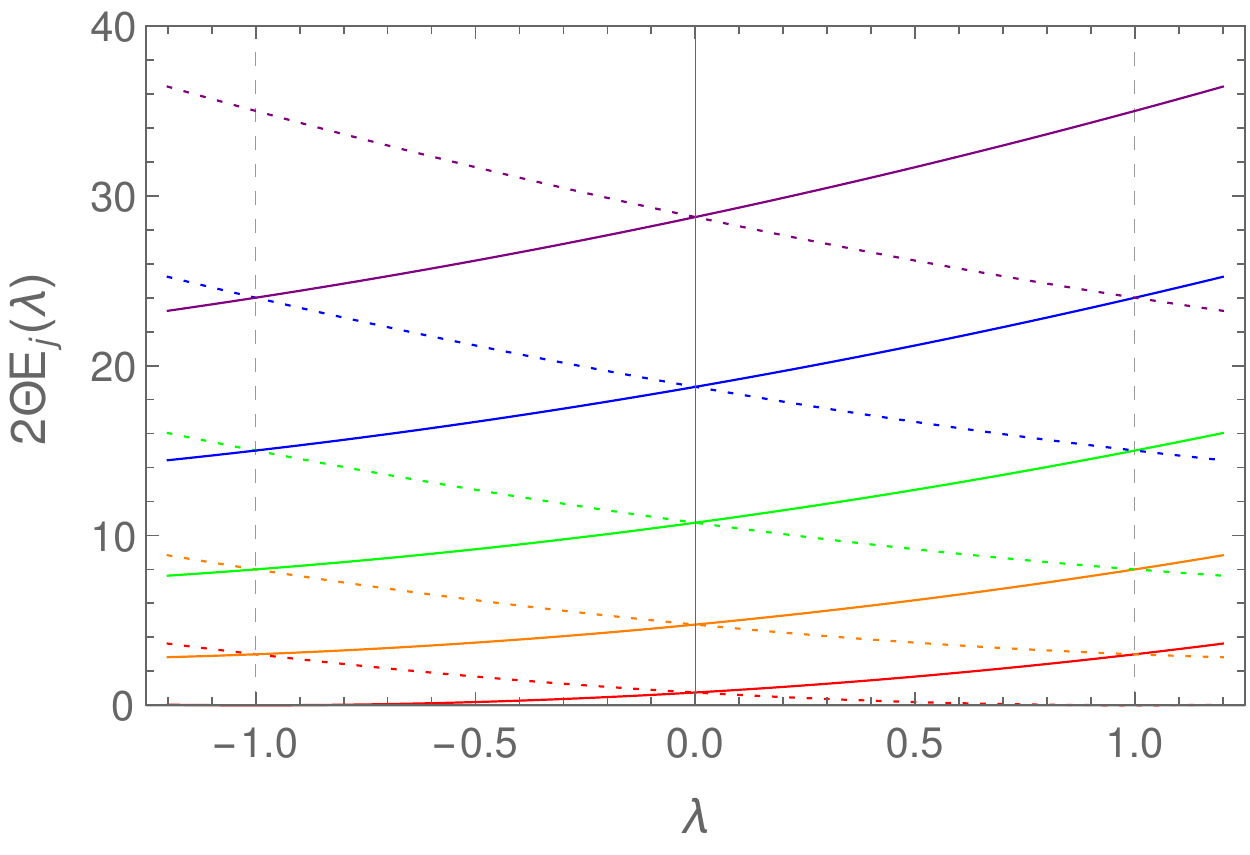}
\caption{\it $\lambda$-dependence of the rotor spectrum in the presence of a
nucleon (with $I = \frac{1}{2}$ and putting $E(\vec p) = 0$). The states, 
which are characterized by $j_L$ and $j_R = j_L - \frac{1}{2}$ (solid lines) 
or $j_R = j_L + \frac{1}{2}$ (dotted lines), are $(2 j_L + 1)(2 j_R + 1)$-fold 
degenerate. The solid and dotted lines that intersect at $\lambda = 0$ have 
the same $j = j_L + j_R \in \{\frac{1}{2}, \frac{3}{2}, \dots\}$, with $j$ 
increasing by 1 as one progresses from one energy $E_j(0)$ to the next.
Interestingly, the solid and dotted lines also intersect at $\lambda = \pm 1$,
now with the corresponding $j$ values differing by 1.}
\label{isospin1_2}
\end{center}
\end{figure}

For $\lambda \neq 0$ the Hamiltonian takes the form
\begin{equation}
H(\lambda) = H(0) + 
\frac{1}{2\Theta} \left(\lambda C + \frac{3}{4} \lambda^2\right) \ ,
\end{equation}
where
\begin{equation}
C = i (\vec \sigma \cdot \vec e_p) \left(\vec e_\alpha \partial_\alpha + 
\frac{1}{\sin\alpha} \vec e_\theta \partial_\theta +
\frac{1}{\sin\alpha \sin\theta} \vec e_\varphi \partial_\varphi -
\tan{\frac{\alpha}{2}} \vec e_\alpha\right) \cdot \vec\tau \ .
\end{equation}
One can convince oneself that $[C,\vec J] = [C,\vec K] = 0$, which shows that
$H(\lambda)$ is still $SU(2)_L \times SU(2)_R$ invariant. One can check explicitly
that
\begin{equation}
C = 2 (\vec \sigma \cdot \vec e_p) \vec J \cdot \vec K \ .
\end{equation}
Finally, one obtains the energy spectrum for $\lambda \neq 0$ as
\begin{equation}
E_j(\lambda) = \frac{1}{2 \Theta} 
\left[j'(j' + 2) + \frac{\lambda^2 - 1}{2}\right] + E(\vec p) \ ,
\end{equation}
where $j' = j \pm \frac{\lambda}{2}$. Now we have two groups of 
$(j + \frac{1}{2})(j + \frac{3}{2})$-fold degenerate states, one for 
$j' = j + \frac{\lambda}{2}$ and the other for $j' = j - \frac {\lambda}{2}$.
The energy spectrum as a function of $\lambda$ is illustrated in 
Fig.\ref{isospin1_2}. 

\section{Rotor Spectrum in the Presence of a Baryon \\
with Arbitrary Isospin}

In this section, we consider baryons of arbitrary isospin, but still with spin 
$\frac{1}{2}$, which includes the $I = 1$ $\Sigma$ baryon in QCD. We extend
the mathematical analysis to arbitrarily large values of the isospin, even if
corresponding baryons are not present in the QCD spectrum. The physical 
$\Sigma$ baryon is stable against strong decays, e.g.\ into a $\Lambda$ baryon
and a pion, which is energetically forbidden. In the chiral limit of massless
up and down quarks (but still with a massive strange quark), on the other
hand, the pions are massless and the decay becomes possible. In a periodic 
volume, the decay process is affected by momentum quantization 
\cite{Wie89,Lue91}. For the moment, we neglect the decay channel 
$\Sigma \rightarrow \Lambda \pi$, and concentrate entirely on how the 
precession of the chiral order parameter is influenced by a baryon of 
arbitrary isospin $I$.

Following the construction in Section 2, for general isospin $I$ the Lagrange 
function takes the form
\begin{equation}
\label{LagrangeI}
L = \frac{3 \Theta}{8 I(I + 1)(2 I + 1)} 
\mbox{Tr}[\partial_t O^\dagger \partial_t O] 
+ \Psi^\dagger \left[E(\vec p) - i \partial_t - i \nu_t - 
i \lambda \vec \sigma \cdot \vec e_p \alpha_t \right] \Psi \ .
\end{equation}
The relation
\begin{equation}
\mbox{Tr}[T_a T_b] = \frac{I(I + 1)(2 I + 1)}{3} \delta_{ab}
\end{equation}
gives rise to the prefactor of the term proportional to $\Theta$, where
\begin{equation}
\frac{I(I + 1)(2 I + 1)}{3} = \sum_{I_3 = -I}^I I_3^2 \ .
\end{equation}

In analogy to the $I = \frac{1}{2}$ case, it is straightforward to derive
a Hamiltonian from the Lagrange function of eq.(\ref{LagrangeI}). The 
result is very simple: the Hamiltonian as well as the generators $\vec L$ 
and $\vec R$ of chiral rotations retain the same form as in the isospin 
$\frac{1}{2}$ case, except that $\frac{1}{2} \vec \tau$ is replaced by the 
corresponding isospin $I$ representation $\vec T$. For $\lambda = 0$ 
the Hamiltonian then takes the form
\begin{equation}
H(0) = \frac{1}{2 \Theta} \left(\vec J^2 + \vec K^2 - I(I + 1)\right) + 
E(\vec p) \ .
\end{equation}
The resulting energy spectrum is thus given by
\begin{equation}
E_{j_L,j_R}(0) = \frac{1}{\Theta} \left(j_L(j_L + 1) + j_R(j_R + 1) - 
\frac{1}{2} I(I + 1)\right) + E(\vec p) \ .
\end{equation}

In the vacuum case, we had $I = 0$ and $j_L = j_R$, while for the nucleon (with
isospin $I = \frac{1}{2}$) we had $j_L = j_R \pm \frac{1}{2}$. For arbitrary
isospin, we have $j_L = j_R + \Delta$, where $\Delta \in \{-I,-I+1,\dots,I\}$. 
These restrictions follow from relations between the two Casimir operators 
$C_1$ and $C_2$. For the vacuum case (with $I = 0$) we had $C_2 = 0$, and for 
the nucleon (with $I = \frac{1}{2}$) we had 
$C_2^2 - \frac{1}{2} C_1 - \frac{3}{16} = 0$. For $I = 1$ the allowed values of 
$\Delta$ are $0$ and $\pm 1$. In this case, the constraint on $\Delta$ follows 
from the relation
\begin{equation}
\label{constraint1}
C_2 (C_2^2 - 2 C_1) = 0 \ ,
\end{equation}
which is possible (but somewhat tedious) to verify explicitly. When $C_2 = 0$ 
(which was the constraint in the $I = 0$ case), one obtains $\Delta = 0$. When 
$C_2^2 - 2 C_1 = 0$, on the other hand, one obtains $\Delta = \pm 1$. In order
to satisfy eq.(\ref{constraint1}), one of these two constraints must be 
satisfied, and hence, for $I = 1$, we indeed obtain $\Delta \in \{-1,0,1\}$. 
Similarly, for $I = \frac{3}{2}$ the two Casimir operators are related by
\begin{equation}
\label{constraint32}
\left(C_2^2 - \frac{1}{2} C_1 - \frac{3}{16}\right)
\left(C_2^2 - \frac{9}{2} C_1 + \frac{45}{16}\right) = 0 \ ,
\end{equation}
which is again non-trivial to verify explicitly. We identify the first bracket 
as the constraint for isospin $\frac{1}{2}$ (which yields 
$\Delta = \pm \frac{1}{2}$), while $\Delta = \pm \frac{3}{2}$ when the second 
bracket vanishes. As a result, for $I = \frac{3}{2}$ eq.(\ref{constraint32}) 
indeed implies 
$\Delta \in \{-\frac{3}{2},-\frac{1}{2},\frac{1}{2},\frac{3}{2}\}$. In the 
$I = 2$ case this story continues and the constraint now takes the form
\begin{equation}
\label{constraint2}
C_2 (C_2^2 - 2 C_1) (C_2^2 - 8 C_1 + 12) = 0 \ .
\end{equation}
We identify the first two factors as the constraints that give rise to
$\Delta = 0,\pm 1$, while $\Delta = \pm 2$ when the third factor vanishes. This
implies that $\Delta \in \{-2,-1,\dots,2\}$ for $I = 2$. Finally, for arbitrary
integer isospin $I$, $\Delta \in \{-I,-I+1,\dots,I\}$ follows from the 
constraint
\begin{equation}
C_2 \prod_{\Delta \in \{1,2,\dots,I\}} [C_2^2 - \Delta^2 (2 C_1 + 1 - \Delta^2)] = 0 
\ ,
\end{equation}
while for half-integer isospin the constraint takes the form
\begin{equation}
\prod_{\Delta \in\{\frac{1}{2},\frac{3}{2},\dots,I\}} 
[C_2^2 - \Delta^2 (2 C_1 + 1 - \Delta^2)] = 0 \ ,
\end{equation}
Again using $j = j_L + j_R$ the corresponding energy spectrum for arbitrary 
isospin is given by
\begin{equation}
E_j(0) = \frac{1}{2 \Theta} \left[j(j+2) + \Delta^2 - I(I+1)\right] + 
E(\vec p) \ .
\end{equation}

\begin{figure}[tbh]
\begin{center}
\includegraphics[width=\columnwidth]{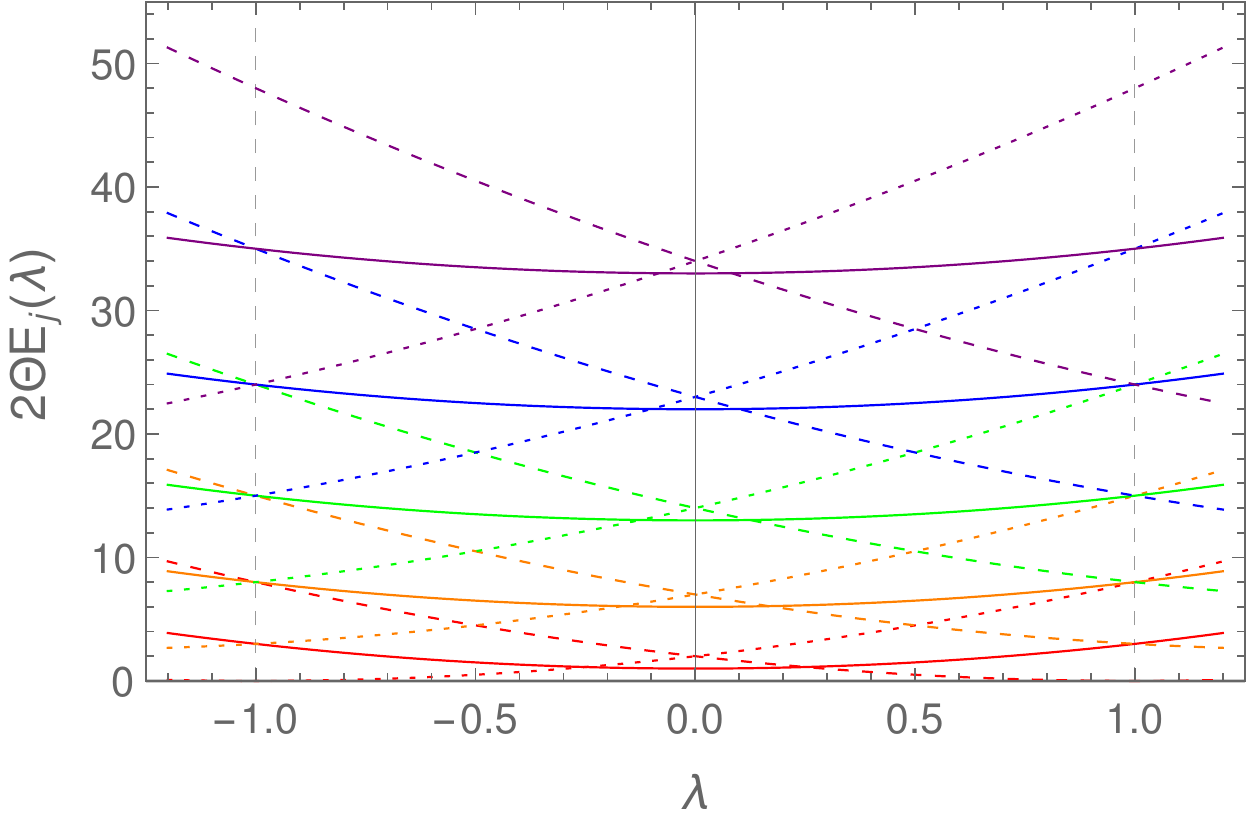}
\caption{\it $\lambda$-dependence of the rotor spectrum in the presence of a
$\Sigma$ baryon (with $I = 1$, putting $E(\vec p) = 0$). The states, which 
are characterized by $j_L$ and $j_R = j_L$ (solid lines), $j_R = j_L - 1$ 
(dotted lines), or $j_R = j_L + 1$ (dashed lines), are 
$(2 j_L + 1)(2 j_R + 1)$-fold degenerate. The dashed and dotted lines that 
intersect at $\lambda = 0$ as well as the solid line below them have the same 
value $j = j_L + j_R \in \{1, 2, \dots\}$, with $j$ increasing by 1 as one 
progresses from one set of three lines to the next. As before, the lines 
intersect again at $\lambda = \pm 1$.}
\label{isospin1}
\end{center}
\end{figure}

\begin{figure}[tbh]
\begin{center}
\includegraphics[width=\columnwidth]{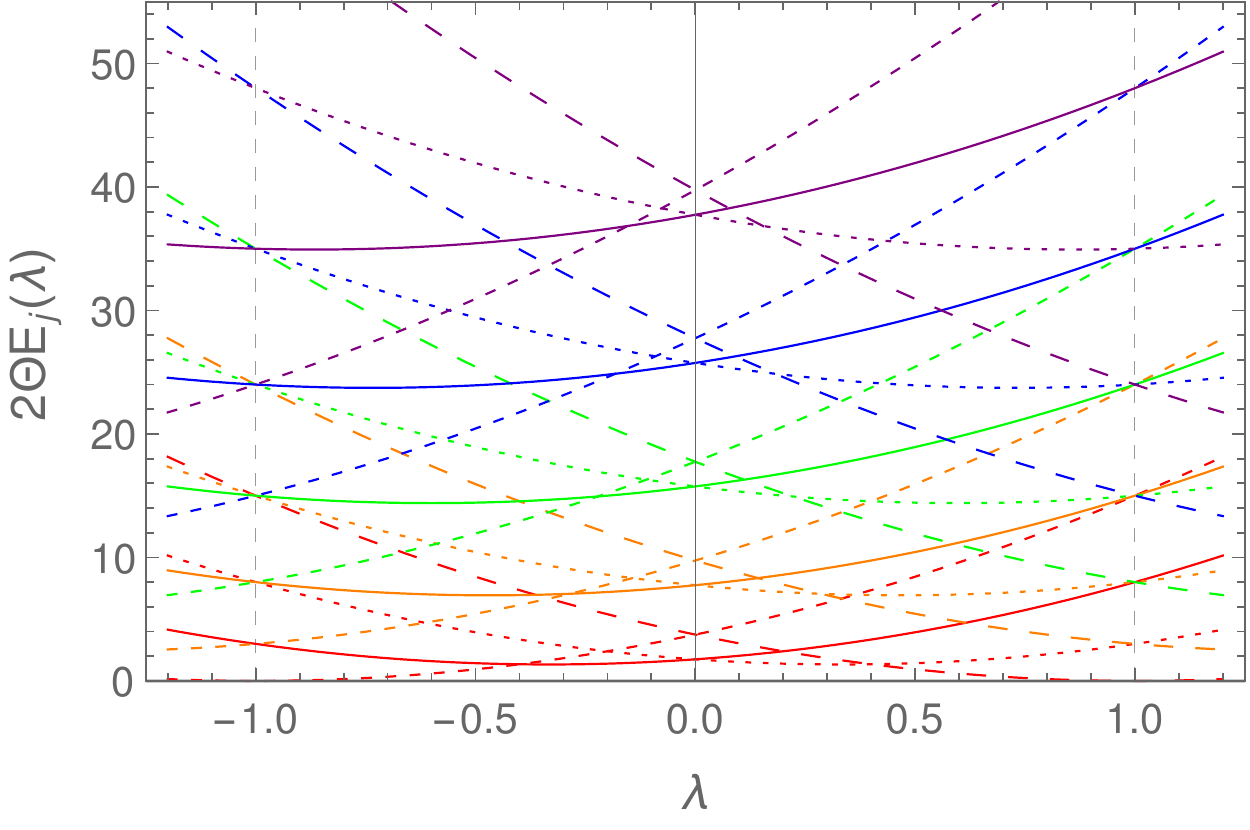}
\caption{\it $\lambda$-dependence of the rotor spectrum in the presence of a
baryon with $I = \frac{3}{2}$ (putting $E(\vec p) = 0$). The states, which are 
characterized by $j_L$ and $j_R = j_L - \frac{3}{2}$ (short-dashed lines), 
$j_R = j_L - \frac{1}{2}$ (solid lines), $j_R = j_L + \frac{1}{2}$ (dotted
lines), or $j_R = j_L + \frac{3}{2}$ (long-dashed lines), are 
$(2 j_L + 1)(2 j_R + 1)$-fold degenerate. The solid and dotted lines that 
intersect at $\lambda = 0$ as well as the short- and long-dashed lines that
intersect above them all have the same value 
$j = j_L + j_R \in \{\frac{3}{2}, \frac{5}{2}, \dots\}$, with $j$ increasing 
by 1 as one progresses from one set of four lines to the next. As before, the 
lines intersect again at $\lambda = \pm 1$.}
\label{isospin3_2}
\end{center}
\end{figure}

For $\lambda \neq 0$ the Hamiltonian takes the form
\begin{equation}
H(\lambda) = H(0) + \frac{1}{2\Theta} \left(\lambda C + \lambda^2 I(I+1)\right) 
\ ,
\end{equation}
and again $C = 2 (\vec \sigma \cdot \vec e_p) \vec J \cdot \vec K$. Hence, the 
energy spectrum now results as
\begin{equation}
E_j(\lambda) = \frac{1}{2 \Theta} \left[j'(j'+2) + 
(\lambda^2 - 1)[I(I+1) - \Delta^2]\right] \ ,
\end{equation}
where $j' = j + \lambda \Delta$. The $\lambda$-dependence of the spectrum is 
illustrated in Fig.\ref{isospin1} for $I = 1$ and in Fig.\ref{isospin3_2} for 
$I = \frac{3}{2}$.

\section{Nature of the Berry Gauge Field}

Let us inspect the Berry gauge field in some detail. First of all, although it 
is not a dynamical field, we investigate its Yang-Mills action
\begin{eqnarray}
&&S_{YM}[A] = - \int_{S^3} d\alpha d\theta d\varphi \sqrt{|g|}  
\frac{1}{4} \mbox{Tr}[F_{ij} F^{ij}] \ , \nonumber \\
&&\sqrt{|g|} = \sin^2\alpha \sin\theta \ ,
\end{eqnarray}
where $|g|$ is the determinant of the metric on the 3-sphere $S^3$ (the $SU(2)$ 
group manifold) with
\begin{equation}
g^{ij} = \mbox{diag}(g^{\alpha\alpha},g^{\theta\theta},g^{\varphi\varphi}) = \mbox{diag}
\left(1,\frac{1}{\sin^2\alpha},\frac{1}{\sin^2\alpha \sin^2\theta}\right) \ .
\end{equation}
We read off the Yang-Mills Lagrange density
\begin{equation}
\label{YangMills}
{\cal L}_{YM} = - \frac{1}{4} \sqrt{|g|} \mbox{Tr}[F_{ij} F^{ij}] = 
- \frac{1}{4} \sqrt{|g|} \mbox{Tr}[F_{ij} g^{ik} g^{jl} F_{kl}] = 
\frac{3}{4} (1 - \lambda^2)^2 \sin^2\alpha \sin\theta \ ,
\end{equation}
which is constant over $S^3$, because it is proportional to the measure factor
$\sin^2\alpha \sin\theta$. The Yang-Mills action of the Berry gauge field then 
takes the value $S_{YM}[A] = \frac{3 \pi^2}{2} (1 - \lambda^2)^2$.

Let us also consider the Chern-Simons action
\begin{eqnarray}
\label{ChernSimons}
&&S_{CS}[A] = - 2 \lambda (\vec \sigma \cdot \vec e_p)
\int_{S^3} d\alpha d\theta d\varphi \sqrt{|g|} \frac{1}{8 \pi^2}
\tilde \varepsilon^{ijk} 
\mbox{Tr}\left[A_i \partial_j A_k + \frac{2}{3} A_i A_j A_k\right], \nonumber \\
&&\tilde \varepsilon^{ijk} = \frac{\varepsilon^{ijk}}{\sqrt{|g|}} \ .
\end{eqnarray}
Here $\tilde \varepsilon^{ijk}$ is the antisymmetric tensor that transforms 
covariantly under general coordinate transformations, while 
$\varepsilon^{ijk} \in \{0,\pm 1\}$ is the ordinary antisymmetric Levi-Civita
symbol. Just as in eq.(\ref{YangMills}), the trace in eq.(\ref{ChernSimons}) 
refers only to isospin but not to spin. Therefore, the expression for the 
Chern-Simons action still involves the matrix-valued prefactor 
$(\vec \sigma \cdot \vec e_p)$. While this may seem strange, it is 
mathematically and physically fully consistent in this context. In particular,
if we quantize the baryon's spin in the direction $\vec e_p$ of its momentum
vector, $(\vec \sigma \cdot \vec e_p)$ reduces to a simple sign $\pm 1$ that
characterizes the baryon's helicity. We now read off the Chern-Simons Lagrange 
density
\begin{eqnarray}
\label{ChernSimonsLagrangian}
{\cal L}_{CS}&=&-  2 \lambda (\vec \sigma \cdot \vec e_p)
\frac{1}{8 \pi^2} \varepsilon^{ijk} 
\mbox{Tr}\left[A_i \partial_j A_k + \frac{2}{3} A_i A_j A_k\right] \nonumber \\
&=&\lambda^2 \frac{1}{8 \pi^2} 
\left((1 - \lambda^2) \sin^2\alpha + 2 \sin^2\frac{\alpha}{2}\right) 
\sin\theta \ .
\end{eqnarray}
Since the Berry vector potential itself also contains the matrix-valued term
$(\vec \sigma \cdot \vec e_p)$, and since $(\vec \sigma \cdot \vec e_p)^2 = \1$,
the actual value of the Chern-Simons term is proportional to the unit-matrix in 
spin space. The value of the Chern-Simons action for the Berry gauge field is 
given by $S_{CS}[A] = \frac{\lambda^2}{4}(3 - \lambda^2)$.

Remarkably, the Berry gauge field solves the Yang-Mills-Chern-Simons classical 
equations of motion on the curved ``space-time'' $S^3$
\begin{eqnarray}
&&D_j \left(\sqrt{|g|} F^{ij}\right) = \partial_j \left(\sqrt{|g|} F^{ij}\right) 
+ \sqrt{|g|} [A_j,F^{ij}] = J^i \ , \nonumber \\
&&J^i = - 2 \lambda  (\vec \sigma \cdot \vec e_p) \varepsilon^{ijk} F_{jk} \ .
\end{eqnarray}
Here $D_j$ is a covariant derivative and $J^i$ is the current induced by the 
Chern-Simons term. The Chern-Simons term itself is not gauge invariant. It 
changes by $2 \lambda (\vec \sigma \cdot \vec e_p)$ times the integer winding 
number 
\begin{eqnarray}
n[\Omega]&=&\frac{1}{24 \pi^2} \int_{S^3} d\alpha d\theta d\varphi \sqrt{|g|} 
\tilde \varepsilon^{ijk} 
\mbox{Tr}\left[(\Omega \partial_i \Omega^\dagger)
(\Omega \partial_j \Omega^\dagger)(\Omega \partial_k \Omega^\dagger)\right] 
\nonumber \\
&=&\frac{1}{24 \pi^2} \int_{S^3} d\alpha d\theta d\varphi \varepsilon^{ijk} 
\mbox{Tr}\left[(\Omega \partial_i \Omega^\dagger)
(\Omega \partial_j \Omega^\dagger)(\Omega \partial_k \Omega^\dagger)\right] 
\nonumber \\
&\in&\Pi_3[S^3] = \Z \ ,
\end{eqnarray}
of the gauge transformation function $\Omega \in SU(2)$. Although the 
Chern-Simons action is not invariant under large gauge transformations, the 
resulting classical equation of motion is gauge covariant. It is important to
point out that, in this context, the prefactor of the Chern-Simons term need 
not be quantized, because the Berry gauge field is not a dynamical quantum 
field. Under a gauge transformation the various fields transform as
\begin{equation}
A_i' = \Omega (A_i + \partial_i) \Omega^\dagger \ , \quad 
F_{ij}' = \Omega F_{ij} \Omega^\dagger \ , \quad
J_i' = \Omega J_i \Omega^\dagger \ .
\end{equation}
The current $J^i$ is covariantly conserved, i.e.
\begin{equation}
D_i J^i = 0 \ ,
\end{equation}
as a consequence of the non-Abelian Bianchi identity
\begin{equation}
\varepsilon^{ijk} D_i F_{jk} = 0 \ ,
\end{equation}
which is automatically satisfied for any non-Abelian field strength
$F_{ij} = \partial_i A_j - \partial_j A_i + [A_i,A_j]$.

While the term proportional to $(\lambda^2 - 1)$ in 
eq.(\ref{ChernSimonsLagrangian}) is 
constant over the 3-sphere, the other term is not, because it is not just
proportional to the measure factor $\sin^2\alpha \sin\theta$. This term, which
is proportional to $\sin^2\frac{\alpha}{2} \sin\theta$, is even singular at 
$\alpha = \pi$, which corresponds to the south-pole of the 3-sphere. Since the 
Yang-Mills Lagrange density of eq.(\ref{YangMills}) is constant, this 
singularity is just a gauge artifact, which can be attributed to a Dirac string 
that passes through the south-pole of $S^3$. The Dirac string emanates
from the origin of $\R^4$ in which we can embed $S^3$. In this sense, the Berry
gauge field configuration is reminiscent of a ``magnetic monopole'' at the
center of the 3-sphere. However, unlike the usual Dirac monopole, this object
lives in 4 instead of 3 ``spatial'' dimensions. In any case, since the Berry 
gauge field is not a physical object in space-time, it does not make too much 
sense to discuss its physical nature as a ``monopole''.  Still, we find it 
remarkable that the Berry gauge field is the solution of a classical equation of
motion on $S^3$. Interestingly, in another context a Berry gauge field has been
identified as the BPS monopole solution of a
Yang-Mills theory coupled to an adjoint Higgs field \cite{Son09}. In that case,
besides the Berry gauge field, the states of the quantum mechanical system also
give rise to an adjoint Higgs field as a ``Berry matter field'' which provides a
covariantly conserved current in the Yang-Mills-Higgs equation of motion. In our
case, instead of ``Berry matter'' the Chern-Simons term of the Berry gauge field
provides a conserved current. While it would be interesting to further 
investigate the Berry gauge field of eq.(\ref{vector_potential}) in a dynamical 
context, for our present purposes the above characterization of its geometrical
and topological features is sufficient.

\section{Conclusions}

We have investigated the rotor spectrum of QCD with two massless and one
massive flavor in a periodic volume in the baryon number 1 sector, for 
different values of the isospin $I = \frac{1}{2}, 1, \frac{3}{2}, 2, \dots$, 
thus generalizing the $I = \frac{1}{2}$ results of \cite{Cha08}. The presence 
of a baryon manifests itself by a Berry gauge field in the effective rotor 
Hamiltonian. Interestingly, the Berry gauge field solves an abstract 
Yang-Mills-Chern-Simons equation of motion in the group space $S^3$ of $SU(2)$.

It would be interesting to further investigate the QCD spectrum in the chiral
limit. In particular, the single-pion states also belong to a tower of rotor
states, presumably again with a non-trivial Berry gauge field. In addition, it
would be interesting to incorporate transitions between baryons of different
isospin, such as $\Sigma \rightarrow \Lambda \pi$, induced by pion emission 
and absorption, and study the corresponding effects on the Berry gauge field. 
Furthermore, one can consider QCD with three massless flavors and investigate
the precession of the chiral order parameter in the corresponding $SU(3)$ 
vacuum manifold. In all these cases, one can ask whether the resulting Berry 
gauge fields again solve a classical equation of motion.

In principle, our analytic results can have an impact on the analysis of lattice
QCD data, which, however, are difficult to obtain in the chiral limit. In order
to make contact with lattice QCD, it would therefore be interesting to extend 
the analytic calculations by including explicit chiral symmetry breaking 
effects due to non-zero up and down quark masses. While our results in the 
strict chiral limit are mostly of academic interest, they shed new light on the
concept of Berry gauge fields and its manifestation in non-trivial quantum field
theories including QCD.

\section*{Acknowledgments}

We like to thank Matthias Blau and Ferenc Niedermayer for illuminating 
discussions. The research leading to these results has received funding from 
the Schweizerischer Na\-tio\-nal\-fonds and from the European Research Council 
under the European Union's Seventh Framework Programme (FP7/2007-2013)/ ERC 
grant agreement 339220.

\end{document}